\documentclass[sigconf]{acmart}
\usepackage{enumitem}
\usepackage{multirow} 
\usepackage{bm}
\usepackage{colortbl}
\usepackage{makecell}
\usepackage{graphicx}
\usepackage{subcaption}
\usepackage{caption}
\usepackage{xspace}

\usepackage[misc]{ifsym} 

\newcommand{\ie}{\emph{i.e.},~}
\newcommand{\eg}{\emph{e.g.},~}
\newcommand{\wrt}{\emph{w.r.t.}~}

\renewcommand{\paragraph}[1]{\medskip\noindent\textbf{#1.~}}

\newcommand{\bmL}{\bm{L}}

\newcommand{\calE}{\mathcal{E}}

\newcommand{\calO}{\mathcal{O}}
\newcommand{\calP}{\mathcal{P}}

\newcommand{\calR}{\mathcal{R}}

\newcommand{\bbI}{\mathbb{I}}

\newcommand{\bbR}{\mathbb{R}}

\newcommand{\modelname}{\textsc{RAT}}
\AtBeginDocument{%
}

\copyrightyear{2024}
\acmYear{2024}
\setcopyright{rightsretained}
\acmConference[WWW '24 Companion]{Companion Proceedings of the ACM Web Conference 2024}{May 13--17, 2024}{Singapore, Singapore}
\acmBooktitle{Companion Proceedings of the ACM Web Conference 2024 (WWW '24 Companion), May 13--17, 2024, Singapore, Singapore}\acmDOI{10.1145/3589335.3651550}
\acmISBN{979-8-4007-0172-6/24/05}

\makeatletter
\gdef\@copyrightpermission{
  \begin{minipage}{0.3\columnwidth}
   \href{https://creativecommons.org/licenses/by/4.0/}{\includegraphics[width=0.90\textwidth]{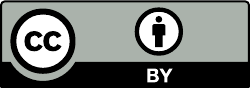}}
  \end{minipage}\hfill
  \begin{minipage}{0.7\columnwidth}
   \href{https://creativecommons.org/licenses/by/4.0/}{This work is licensed under a Creative Commons Attribution International 4.0 License.}
  \end{minipage}
  \vspace{5pt}
}
\makeatother

\title[\modelname{}: Retrieval-Augmented Transformer for Click-Through Rate Prediction]{\modelname{}: Retrieval-Augmented Transformer\\ for Click-Through Rate Prediction}

\author{Yushen Li}
\authornote{Both authors contributed equally to this work.}
\author{Jinpeng Wang}
\authornotemark[1]
\affiliation{%
  \institution{Tsinghua Shenzhen International Graduate School, Tsinghua University}
  \city{Shenzhen}
  \country{China}
}
\email{liyushen22@mails.tsinghua.edu.cn}
\email{wjp20@mails.tsinghua.edu.cn}

\author{Tao Dai}
\affiliation{%
  \institution{College of Computer Science and Software Engineering, Shenzhen University}
  \city{Shenzhen}
  \country{China}
}
\email{daitao.edu@gmail.com}

\author{Jieming Zhu}
\authornote{Corresponding authors.}
\affiliation{%
  \institution{Huawei Noah’s Ark Lab}
  \city{Shenzhen}
  \country{China}
}
\email{jiemingzhu@ieee.org}

\author{Jun Yuan}
\affiliation{%
  \institution{Huawei Noah’s Ark Lab}
  \city{Shenzhen}
  \country{China}
}
\email{yuanjunfy@163.com}

\author{Rui Zhang}
\affiliation{%
  \institution{www.ruizhang.info}
  \city{Shenzhen}
  \country{China}
}
\email{rayteam@yeah.net}

\author{Shu-Tao Xia}
\authornotemark[2]
\affiliation{%
  \institution{Tsinghua Shenzhen International Graduate School, Tsinghua University}
  \city{Shenzhen}
  \country{China}
}
\affiliation{%
  \institution{Research Center of Artificial Intelligence, Peng Cheng Laboratory}
  \city{Shenzhen}
  \country{China}
}
\email{xiast@sz.tsinghua.edu.cn}

\renewcommand{\shortauthors}{Li and Wang, et al.}
\begin{abstract}
Predicting click-through rates (CTR) is a fundamental task for Web applications, where a key issue is to devise effective models for feature interactions. 
Current methodologies predominantly concentrate on modeling feature interactions within an individual sample, while overlooking the potential cross-sample relationships that can serve as a reference context to enhance the prediction. 
To make up for such deficiency, this paper develops a \underline{\textbf{R}}etrieval-\underline{\textbf{A}}ugmented \underline{\textbf{T}}ransformer (\modelname{}), aiming to acquire \emph{fine-grained} feature interactions \emph{within} and \emph{across} samples. 
By retrieving similar samples, we construct augmented input for each target sample. 
We then build Transformer layers with cascaded attention to capture both intra- and cross-sample feature interactions, facilitating comprehensive reasoning for improved CTR prediction while retaining efficiency.  
Extensive experiments on real-world datasets substantiate the effectiveness of \modelname{} and suggest its advantage in long-tail scenarios.
The code has been open-sourced at \texttt{\textcolor{purple}{\url{https://github.com/YushenLi807/WWW24-RAT}}}.
\end{abstract}



\begin{document}
\maketitle
\section{Introduction}
\label{sec: introduction}

\begin{figure}[!t]
    \centering
    \includegraphics[width=0.47\textwidth]{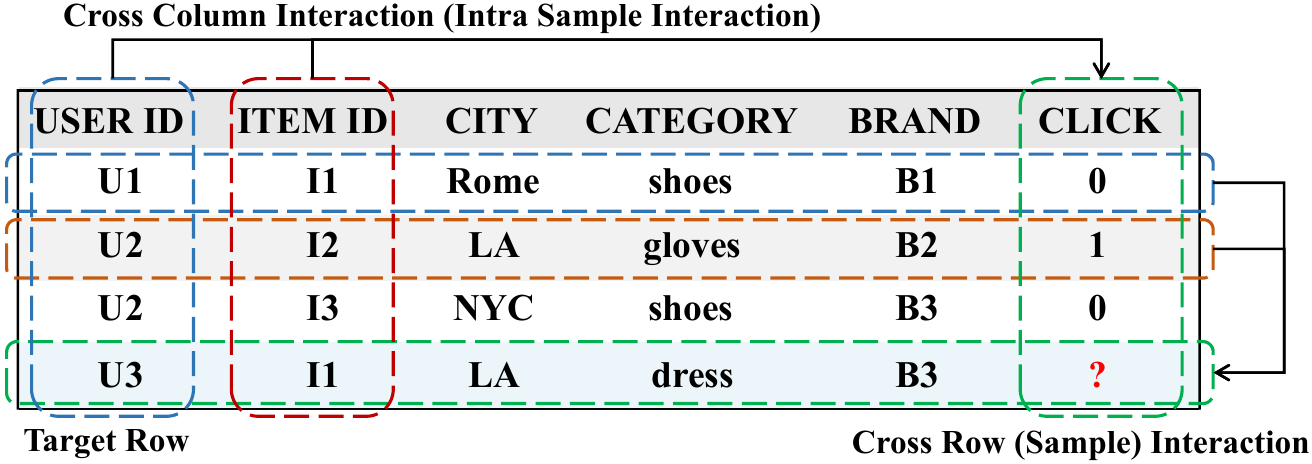}
    \caption{Traditional CTR prediction methods only focus on cross-column interaction.
    In this paper, we focus on a new paradigm of retrieval-augmented (RA) CTR, further incorporating cross-row interaction for more effective prediction.}
    \label{fig:compare}
\end{figure}

Click-through rate (CTR) prediction is a binary classification task that aims to forecast whether a user will click on a given item.
It has been broadly applicable in commercial fields such as advertising placement and recommender systems \cite{xdeepfm,deepfm,wang2023missrec,refer}. 
Feature interaction modeling plays an essential role in CTR prediction.  
As shown in Figure \ref{fig:compare}, traditional methods \cite{interhat,finalmlp,wdl,fm} primarily focus on feature interactions within each sample, but seldom consider cross-sample information that can serve as a reference context to enhance the prediction. 
Since features and their interactions are usually sparse, it necessitates CTR models to capture and memorize all interaction patterns, posing challenges in robustness and scalability. 

Recently, retrieval-augmented (RA) learning has shown effective in natural language processing~\cite{realm} and computer vision~\cite{chen2022re}, whose typical idea is to retrieve similar samples and enhance model prediction with these external demonstrations. 
Inspired by its success in relieving long-tail problems~\cite{long2022retrieval}, we believe it is a promising paradigm to relieve the aforementioned issue in CTR prediction. 
In this direction, RIM  \cite{rim}, DERT \cite{dert} and PET \cite{pet} are three preliminary works on RA CTR prediction. 
However, they either compromise intra- or cross-sample feature interaction, which are still sub-optimal practices. 
Specifically, RIM simply aggregates retrieved samples on each feature field, which sacrifices fine-grained cross-sample knowledge. 
Although DERT improves RIM on sample retrieval and supplies better reference samples to the target sample, it still follows the same retrieval-augmented modeling as RIM, thereby sharing common limitations.
PET constructs a hypergraph to represent the relations between rows and columns of tabular data. 
Via the star expansion, it converts the hypergraph into a heterogeneous graph that includes label nodes and feature nodes. 
PET can effectively capture the cross-sample information by designing a message-passing mechanism, but all feature nodes need to be transitioned through label nodes on the graph, showing inflexibility in modeling intra-sample feature interactions. 

To remedy the shortcomings in previous works, we propose a unified framework termed \underline{\textbf{R}}etrieval-\underline{\textbf{A}}ugmented \underline{\textbf{T}}ransformer (\modelname{}) to enhance fine-grained intra- and cross-sample feature interactions for CTR prediction. 
Given a target sample, we retrieve similar samples from a reference pool (\eg historical logs) using the sparse retrieval algorithm. 
Then we develop a Transformer-based model to acquire fine-grained feature interactions within and across samples. 
In particular, we find that intra-cross cascade attention not only improves the efficiency beyond joint modeling but also enhances the robustness of \modelname{}. 
Without bells and whistles, we condense the semantic information to one token representation, which is fed to the binary classifier to make the final prediction. 

We conduct extensive experiments on three real-world datasets: ML-Tag, KKBox, and Tmall, demonstrating the promise of retrieval-augmented approaches and the further improvement of \modelname{}. 
We also show that \modelname{} can enhance long-tail sample prediction, which suggests its capacity to tackle feature sparsity and cold start issues.

To summarize, we make the following contributions:
\setlist{nolistsep}
\begin{itemize}[leftmargin=1.5em]
\item[$\clubsuit$] We propose a Retrieval-augmented Transformer (\modelname{}) for CTR prediction, which enhances fine-grained intra- and cross-sample feature interaction in a unified model. 
\item[$\clubsuit$] We find that intra-cross cascade attention not only improves the efficiency but also enhances the robustness of \modelname{} (\S\ref{subsubsec:block_design}). 
\item[$\clubsuit$] Extensive experiments on real-world datasets validate \modelname{}'s efficacy and suggest its advantage against the feature sparsity and cold start issues. 
\end{itemize}

\section{The Proposed Method}
\label{sec:method}
We design \textbf{RE}trieval-augmented \textbf{T}ransformer (\modelname{}), considering both intra- and cross-sample interactions for CTR prediction. 
Figure \ref{fig:framework} briefly illustrates the framework of \modelname{}. 

\subsection{Retrieve Similar Samples as Context}
Given the $F$-field record $x_i=[x_i^1;...;x_i^F]$ of a target sample, we search for similar samples as the reference context from a reserved sample pool $\calP$. 
We use BM25 \cite{bm25} for retrieval because of its training-free nature,
which also aligns with previous works \cite{rim,pet}. 
Specifically, the relevant score of query $x_i$ and the key $x_c$ of a candidate sample $(x_c,y_c)\in\calP$ is defined as
\begin{gather}
    s(x_t,x_c)=\sum\limits_{f=1}^F\log\frac{N_\calP-N_\calP(x^f_t)+0.5}{N_\calP(x_t^f)+0.5}\cdot\bbI_{\{x_t^f=x_c^f\}},
\end{gather}
where $ \bbI_{\{\cdot\}}$ is an indicator function. 
$N_\calP$ is the number of samples in $\calP$, while $N_\calP(x_t^f)$ denotes the number of samples containing the feature $x_t^f$ in $\calP$. 
Different from \citet{rim,pet} that implemented the retrieval with Elasticsearch on CPUs, we provide an efficient GPU-based implementation to enable faster speed. 

Finally, we retrieve $K$ samples from $\calP$ with the highest scores:
\begin{equation}
    \calR_i=\{(x_{c_1}, y_{c_1}), (x_{c_2}, y_{c_2}), \cdots, (x_{c_K}, y_{c_K})\}.
\end{equation}

\paragraph{Notes on Avoiding Information Leakage} 
We sort the samples in chronological order if there is timestamp information and restrict a query only to retrieve samples that occur earlier than it. 
For validation and testing, we take the whole training set as the reference pool. This strategy still satisfies the restriction and is safe, because the testing set and the validation set are the latest and the next latest parts of the whole dataset in our experiments. 

\subsection{Construct Retrieval-augmented Input}
We build an embedding layer to transform discrete features into $D$-dimensional embedding vectors. 
In particular, we treat the labels of retrieved samples as special features and also build an embedding table for that field. 
Let us denote the set of embedding tables as $\calE=\{\bm{E^1}, \bm{E^2}, \cdots, \bm{E^F}, \bm{L}\}$, where $\bm{E^f}$ is the embedding table of the $f$-th feature field. 
$\bmL\in\bbR^{3\times D}$ is the label embedding table, where the $0$-th and the $1$-st slots are for unclick and click labels of retrieved samples, respectively, and the $2$-nd slot is reserved for the target sample to represent the unknown click label (\ie \texttt{[UNK]}). 

For a retrieved sample $(x_{c_k},y_{c_k})$, we lookup feature embeddings and the label embedding from the embedding tables and obtain $E_{c_k}=[l_{c_k};e_{c_k}^1;e_{c_k}^2;\cdots;e_{c_k}^F]\in\bbR^{(F+1)\times D}$. 
Analogously, for the target record $x_i$, we obtain $E_i=[l_i^\texttt{[UNK]};e_i^1;e_i^2;\cdots;e_i^F]\in\bbR^{(F+1)\times D}$. 
Finally, the retrieval-augmented input for the target record $x_i$ is obtained by stacking $E_i$ and $E_{c_1},E_{c_2},\cdots,E_{c_K}$ together:
\begin{equation}
    \tilde{E}_i=[E_i;E_{c_1};E_{c_2};\cdots;E_{c_K}]\in\bbR^{(K+1) \times (F+1) \times D}.
\end{equation}

\begin{figure}[!t]
    \centering
    \includegraphics[width=0.47\textwidth]{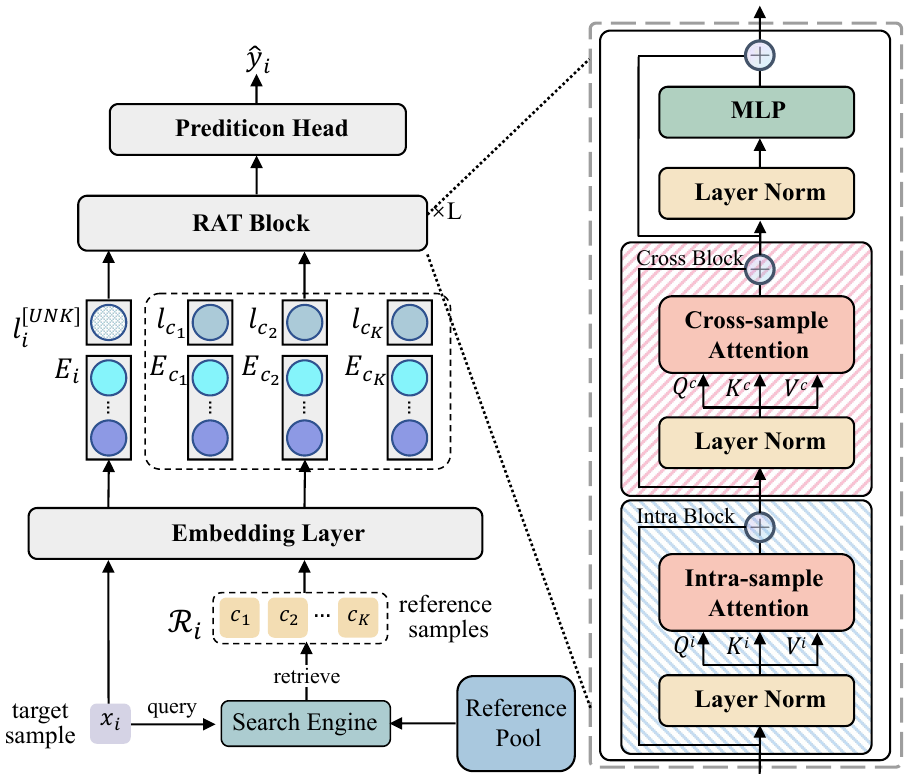}
    \caption{The overview framework of \modelname{}.}
    \label{fig:framework}
\end{figure}

\subsection{Intra- and Cross-sample Feature Interaction}
To integrate intra- and cross-sample feature interactions to enhance CTR prediction, a na\"ive idea is to unstack all retrieved samples, append them to the target record as extra feature fields, and use joint attention to model full 
feature interactions. 
However, it exhibits an efficiency issue: the complexity of each joint self-attention is $\calO((K+1)^2\cdot(F+1)^2)$. 
We also find it inferior in performance (Table \ref{ablation_performance}), possibly due to the influence of noisy feature interactions. 

To address the above issues, we decompose the joint attention. 
As shown in Figure \ref{fig:framework}, each \modelname{} block comprises the cascade of an intra-block, a cross-block, and a multi-layer perception (MLP). 
The forward process of the $\ell$-th \modelname{} block is formulated as 
\begin{gather}
    H_i^\ell=\mathrm{ISA}(\mathrm{LN}(X_i^\ell))+X_i^\ell, \\
    H_i^{'\ell}=\mathrm{CSA}(\mathrm{LN}(H_i^\ell))+H_i^\ell,  
    \label{eq20}\\
    X_i^{\ell+1}=\mathrm{MLP}(\mathrm{LN}(H_i^{'\ell}))+H_i^{'\ell},  
\end{gather}
where $X^\ell_i$ denote the input of the $\ell$-th \modelname{} block. 
$X^0_i=\tilde{E}_i$. 
$H_i^\ell$ and $H_i^{'\ell}$ denote the hidden states. 
$\mathrm{LN}(\cdot)$ is the layer norm operation. 
$\mathrm{ISA}(\cdot)$ and $\mathrm{CSA}(\cdot)$ represent the intra-sample and cross-sample attention modules, respectively, which perform multi-head self-attentions along the field-axis and sample-axis, respectively.

Compared with the vanilla joint attention, our design of cascaded attention favorably reduces the complexity to $\mathcal{O}((K+1)^2 + (F+1)^2)$, which helps to keep efficient. 
We will also investigate more block designs in experiments (\S\ref{subsubsec:block_design}) and demonstrate the advantages of cascaded attention for \modelname{}.
\section{Experiments}
\label{sec:experiments}

\begin{table}[]
\caption{Statistics of datasets}
\resizebox{.9\columnwidth}{!}{
\begin{tabular}{ccccc}
\hline\hline
Dataset    & \#Samples       & \#Fields     & Missing\_ratio  & Positive\_ratio  \\ \hline
ML-tag      &  2,006,859              &    3         & 0\%  & 33.33\%         \\
KKBox      &  7,377,418              &    19         & 6.08\%  & 50.35\%         \\
Tmall       &   54,925,331             &   9       &0.36\%  & 50\%           \\
\hline \hline

\end{tabular}
}
\label{statistics}
\end{table}

\subsection{Experimental Setup}
\label{subsec:setup}

\subsubsection{Datasets}. We evaluate \modelname{} on three popular datasets. \textbf{ML-tag} \cite{ml1},  \textbf{KKBox} \cite{bars} and \textbf{Tmall} \cite{pet}.  We follow the standard CTR prediction benchmark, BARS \cite{bars}, to split ML-tag to train, validation and test sets by 7:2:1,
and split KKBox by 8:1:1. For Tmall, we follow \cite{rim} by dividing the data based on global timestamps to avoid information leakage. Dataset statistics is shown in Table \ref{statistics}.

\subsubsection{Metrics} We use two commonly used metrics, \textbf{AUC} (Area Under ROC) and \textbf{Logloss} (cross entropy) as the evaluation metrics.

\subsubsection{Baselines} 
We select two types of baselines. 
(\textbf{i}) Traditional (and not retrieval-augmented) models: DeepFM \cite{deepfm}, xDeepFM \cite{xdeepfm}, DCNv2 \cite{dcnv2}, AOANet \cite{aoanet}; 
(\textbf{ii}) Retrieval-augmented models: RIM \cite{rim}, PET \cite{pet}. 
For fair comparison, we use BM25 for aligning settings. 
DERT \cite{dert} is not included in our benchmark because it is a special case of RIM with a new dense retrieval setting. 
Besides, there is no open-sourced implementation for it, making fair comparison hard.

\subsubsection{Implementation Details} For fair comparisons, we implement all methods with FuxiCTR \cite{FuxiCTR} and follow the BARS \cite{bars} benchmark settings. For retrieval-augmented models, we set $K$ to 5 by default.

\subsection{Comparison with State-of-the-arts}
\subsubsection{Overall Performance}
We report the model performances in Table \ref{performance}. 
Note that for CTR prediction, a \textperthousand-level AUC increase is considered acceptable, as even such a minor improvement, if statistically significant, can lead to substantial gains in revenue \cite{xdeepfm,fibinet}. 
According to the results in Table \ref{performance}, retrieval-augmented (RA) models generally outperform traditional models, indicating that incorporating cross-sample information 
is effective in enhancing CTR prediction. 
Although PET does not surpass xDeepFM in overall performance, mainly due to its insufficient capability to capture intra-sample feature interactions, it exhibits superiority over other traditional models. 
Moreover, \modelname{} outperforms the SoTA RA model, RIM, demonstrating the strong capability of Transformer in modeling fine-grained intra- and cross-sample interactions.

\begin{table}[]
\caption{CTR prediction performance comparison. $\Delta AUC$ indicates averaged AUC improvement compared to xDeepFM.} 
\resizebox{1\columnwidth}{!}{
\begin{tabular}{c|cc|cc|cc|c}
\hline\hline
\multirow{2}{*}{Model}                  & \multicolumn{2}{c|}{ML-tag}                          & \multicolumn{2}{c|}{KKbox}        & \multicolumn{2}{c|}{Tmall}        & \multirow{2}{*}{$\Delta AUC\uparrow$}    \\ \cline{2-7} 
                                & AUC$\uparrow$                           & Logloss$\downarrow$                       & AUC$\uparrow$             & Logloss$\downarrow$         & AUC$\uparrow$             & Logloss$\downarrow$         &        \\ \hline
DeepFM                               & 0.9685 & 0.2130                        & 0.8429          & 0.4895          & 0.9401          & 0.3341          & -0.24\% \\
DCNv2                                & 0.9691 & 0.2147 & 0.8303          & 0.5073          & 0.9391          & 0.3377          & -0.75\% \\
AOANet                              & 0.9694 &	0.2105& 	0.8313 &	0.5046& 	0.9391& 	0.3369           & -0.70\% \\
xDeepFM                              &  0.9697 &  0.2409& 0.8475          & 0.4846          & 0.9406          & 0.3513          & 0.00\%  \\ \hline
PET                                  & 0.9692                        & 0.2602                        & 0.8316          & 0.5044          & 0.9520          & 0.3175 & -0.24\% \\
RIM                                  & 0.9711                        & 0.1832                        & 0.8465          & 0.4907          & 0.9534          & 0.3131          & 0.46\%  \\ 
\rowcolor{green!10}\textbf{\modelname{}} & \textbf{0.9809}               & \textbf{0.1421}               & \textbf{0.8500} & \textbf{0.4812} & \textbf{0.9589} & \textbf{0.3091}          & \textbf{1.13\%}  \\
\hline\hline
\end{tabular}
}
\label{performance}
\end{table}

\subsubsection{Performance on Long-tail Data}
\begin{table}[t]
\caption{Performance \wrt long-tail users on ML-tag subset.}
\resizebox{.75\columnwidth}{!}{
\begin{tabular}{cc|cc|cc}
\hline\hline
\multirow{2}{*}{Type} & \multirow{2}{*}{Model}  & \multicolumn{2}{c|}{Tail 10\% Users}  & \multicolumn{2}{c}{Tail 20\% Users}    \\ \cline{3-6} 
          &             & AUC$\uparrow$             & Logloss$\downarrow$ &        AUC$\uparrow$             & Logloss$\downarrow$         \\ \hline
\multirow{4}{*}{Non-RA} & DeepFM                 & 0.9487          & 0.2714       & 0.9616          & 0.2319       \\
& DCNv2                                          & 0.9445 &	0.2748 &   0.9578  &	0.2388 \\
& AOANet                                         &0.9479 	&0.2764 &	0.9617 	&0.2344  \\ 
& xDeepFM                                        & 0.9482          & 0.2866  &     0.9611          & 0.2436          \\ \hline
\multirow{3}{*}{RA} & PET                        
& 0.9496          & 0.2746 &        0.9631          & 0.2296
\\
& RIM                                            & 0.9543          & 0.2574  &       0.9654          & 0.2200         \\
  &  \cellcolor{green!10}\textbf{\modelname{}} &  \cellcolor{green!10}\textbf{0.9583} & \cellcolor{green!10}\textbf{0.2250 } &  \cellcolor{green!10}\textbf{0.9727} & \cellcolor{green!10}\textbf{0.1799 } \\ 
\hline\hline
\end{tabular}
}
\label{long_effect}
\end{table}

Inspired by the success of RA mechanisms on long-tail recognition \cite{long2022retrieval}, here we pose a question: Can \modelname{} enhance CTR prediction on long-tail data? 
Taking ML-tag for a case study, we investigate the model efficacy on sample subsets associated with tailed 10\% and 20\% users, respectively, ranked by user's sample amount. 
As shown in Table \ref{long_effect}, RA models perform better than traditional ones, indicating RA context indeed assists reasoning over sparse features and their interaction. 
Furthermore, \modelname{} outperforms RIM and PET considerably, validating that Transformer-based \emph{fine-grained} modeling improves the utilization of retrieved samples. 
The evidence suggests \modelname{}'s capability in addressing important issues like feature sparsity and cold start.

\subsection{Model Analysis}
\subsubsection{Designs of \modelname{} Block} \label{subsubsec:block_design}
\begin{table}[t]
\caption{Performance comparison of different \modelname{} designs.} 
\vspace{-.5em}
\resizebox{1\columnwidth}{!}{
\begin{tabular}{c|cc|cc|cc}
\hline\hline
\multirow{2}{*}{Model}                  & \multicolumn{2}{c|}{ML-tag}                          & \multicolumn{2}{c|}{KKbox}        & \multicolumn{2}{c}{Tmall}          \\ \cline{2-7} 
                                & AUC$\uparrow$                           & Logloss$\downarrow$                       & AUC$\uparrow$             & Logloss$\downarrow$         & AUC$\uparrow$             & Logloss$\downarrow$               \\ \hline
\modelname{}$_{\texttt{JM}}$ & 0.9667                        & 0.2003                        & 0.8415          & 0.4917          & 0.9581          & 0.3110        \\
\modelname{}$_{\texttt{CE}}$ & 0.9736                        & 0.1731                        & 0.8483          & 0.4831          & 0.9575          & 0.3182          \\
\modelname{}$_{\texttt{PA}}$ & 0.9777                        & 0.1557                        & 0.8484          & 0.4828          & 0.9582          & 0.3177      \\ 
\rowcolor{green!10}\textbf{\modelname{}} & \textbf{0.9809}               & \textbf{0.1421}               & \textbf{0.8500} & \textbf{0.4812} & \textbf{0.9589} & \textbf{0.3091}           \\ \hline\hline
\end{tabular}
}
\label{ablation_performance}
\end{table}

\begin{table}[]
\caption{Efficiency comparison of different \modelname{} designs.}
\vspace{-.5em}
\resizebox{1\columnwidth}{!}{
\begin{tabular}{c|cc|cc|cc}
\hline\hline
\multirow{3}{*}{Model} & \multicolumn{2}{c|}{ML-tag} & \multicolumn{2}{c|}{KKBox} & \multicolumn{2}{c}{Tmall} \\
\cline{2-7}
                       & \makecell[c]{Params \\ $\times 10^6\!\!\downarrow$ }     & \makecell[c]{Runtime \\ ($\upmu$s)$\downarrow$}      & \makecell[c]{Params\\ $\times 10^6\!\!\downarrow$ }       & \makecell[c]{Runtime \\ ($\upmu$s)$\downarrow$} & \makecell[c]{Params\\ $\times 10^6\!\!\downarrow$ }       & \makecell[c]{Runtime \\ ($\upmu$s)$\downarrow$}        \\ \hline
\modelname{}$_\texttt{JM}$	&	0.34	&	10.31 	               &	0.46	&	15.18 &	16.87&	12.60 \\ 
\modelname{}$_\texttt{CE}$	&	0.35	&	9.88 	                  &	0.47	&	12.70 &	16.88&	9.91 \\	
\modelname{}$_\texttt{PA}$	&	0.34	&	9.60 	                  &	0.46	&	11.70	&16.87	& 9.78 \\	
\rowcolor{green!10}\modelname{}	&	0.34	&	9.65 	         &	0.46	&	11.76 	&	16.87 & 9.72 \\	
\hline\hline  
\end{tabular}
}
\label{ablation_eff}
\end{table}

In addition to the proposed design, we further explore more variants:
(\textbf{i}) \textbf{Intra-Cross \texttt{J}oint \texttt{M}odeling (\texttt{JM})}: The vanilla joint attention modeling for all intra- and cross-sample feature interactions.
(\textbf{ii}) \textbf{Intra-Cross \texttt{C}ascaded \texttt{E}ncoder (\texttt{CE})}: 
Rather than the cascaded attentions in a single block, it separates the intra- and cross-sample attention into two cascaded Transformer blocks, yielding a double block number compared to other variants.
(\textbf{iii}) \textbf{Intra-Cross \texttt{P}arallel \texttt{A}ttention (\texttt{PA})}: 
It designs intra- and cross-sample attentions as two parallel branches. 
The hidden dimension of each branch is halved, and the outputs of the two branches are concatenated. 
We report the experimental results on model performance and efficiency in Tables \ref{ablation_performance} and \ref{ablation_eff}, respectively.
From Table \ref{ablation_eff} we can see that decomposed modeling designs are indeed more effective than joint modeling (\ie \texttt{JM}). 
From Table \ref{ablation_performance} we can learn that joint modeling is not the best practice, possibly due to the influence of noisy feature interactions. 
In contrast, attention decomposition provides an inductive bias for model robustness and leads to better performance in general. 
Comparing \modelname{} and \modelname{}$_\texttt{CE}$, block-level decomposition does not bring extra gain upon attention-level decomposition. 
Comparing \modelname{} and \modelname{}$_\texttt{PA}$, cascaded attention performs slightly better, suggesting that the successive order benefits the feature interaction modeling. 

\section{Conclusion}
\label{sec:conclusion}
In this paper, we present \underline{\textbf{R}}etrieval-\underline{\textbf{A}}ugmented \underline{\textbf{T}}ransformer (\modelname{}) for CTR prediction.
It highlights the importance of cross-sample information and explores Transformer-based architectures to capture \emph{fine-grained} feature interactions \emph{within} and \emph{between} samples, effectively remedying the shortcomings of existing works. 
By decomposing the intra- and cross-sample interaction modeling, \modelname{} enjoys better efficiency while further enhancing the robustness. 
We conduct comprehensive experiments to validate the effectiveness of \modelname{}, and show its advantage in tackling long-tail data. 

\begin{acks}
This work is supported in part by the National Natural Science Foundation of China under Grant (62302309, 62171248), Shenzhen Science and Technology Program (JCYJ20220818101012025), and the PCNL KEY project (PCL2023AS6-1). 
We gratefully acknowledge the support of MindSpore\footnote{https://www.mindspore.cn}, which is a new deep learning framework for this research.
\end{acks}
\bibliographystyle{ACM-Reference-Format}
\bibliography{main}

\end{document}